\documentclass[prd,aps,floats,tabularx,preprintnumbers,twocolumn]{revtex4}
\usepackage{graphicx, epsfig}

\newcommand{\be}{\begin{equation}}
\newcommand{\ee}{\end{equation}}
\newcommand{\bea}{\begin{eqnarray}}
\newcommand{\eea}{\end{eqnarray}}

\begin{document}

\preprint{SU-GP-02/11-1}
\preprint{SU-4252-771}

\title{Are Domain Walls ruled out ?}

\author{Luca Conversi$^\flat$, Alessandro Melchiorri$^{\flat^,+}$,
Laura Mersini$^*$ and Joseph Silk$^\sharp$}
\affiliation{
$^\flat$ Physics Department, University of Rome
``La Sapienza'', Ple Aldo Moro 2, 00185, Rome, Italy\\
$^+$ Istituto Nazionale di Fisica Nucleare - Sezione di Roma
 - Università degli Studi di Roma "La Sapienza"
 P.le Aldo Moro, 2, 00185 Rome, Italy\\
$^*$ Department of Physics and Astronomy, UNC-Chapel Hill, Phillips Hall,CB 3255,Chapel Hill, NC  27599-3255,USA.\\
$^\sharp$ Astrophysics, Denys Wilkinson Building, University of Oxford, Keble
road, OX1 3RH, Oxford, UK\\
}
\begin{abstract}

Recent analysis of the combined data of cosmic microwave background, galaxy
clustering and supernovae type Ia observations have set strong constraints 
on the equation of state parameter $w_X$. The upper bound $w_X < -0.82$ at $95
\%$ c.l. rules out an important class of models, the domain walls 
($-2/3 < w_X < -1/3$). Here we revisit the issue of  domain walls as a possible
alternative to the
standard $\Lambda$-CDM model by questioning the assumptions made in the
choice of priors of the data analysis. The results of our investigation
show that domain walls can provide a good fit
to the WMAP data for a different choice of priors with ``lower'' values of
the Hubble parameter ($h<0.65$),
(as indicated by Sunyaev-Zeldovich and time delays for gravitational lensing
observations), and ``higher'' values of the matter density
($\Omega_m > 0.35$), (in agreement with recent measurements of the
temperature-luminosity relation of distant clusters observed with
the XMM-Newton satellite). In this new perspective, their existence would 
lead to important implications  for the CMB constraints on cosmological and 
inflationary parameters.

\end{abstract}


\maketitle

\section{Introduction.}

The recent results of precision cosmology and the measurements of Cosmic
Microwave Background Anisotropies have been exremely important since 
they provide an excellent agreement of our theoretical picture of the cosmos,
incorporating the standard model of structure formation, the inflationary 
prediction of flatness, the presence of
cold dark matter and an amount of baryonic matter consistent
with Big Bang Nucleosynthesis constraints (see e.g.
~\cite{spergel},  ~\cite{netterfield}).
The price-tag of this success story of the combined observations of CMB
with   complementary cosmological data concerns a very puzzling consequence: 
the evolution of the universe
is dominated by a mysterious form of energy, $X$, coined dark energy,
(an unclestered negative pressure component of the mass-energy density), 
with a present-day energy density fraction $\Omega_X \simeq 2/3$
and equation of state $w_X \sim -1$ (see e.g. ~\cite{tegmark},
~\cite{spergel}, ~\cite{mmot}). This discovery may turn out to be 
one of the most important contribution to physics in our generation.
Hence it is especially important to consider all possible scheme
for dark energy.

A true cosmological constant
$\Lambda$ may be at works here. Hence it is entirely possible that
a dynamic mechanism is giving rise to the observed acceleration of the
present Universe.
Some of the popular proposed candidates
to explain the observations are a slowly-rolling scalar field,
``quintessence''~\cite{Wetterich:fm}-\cite{Caldwell:1997ii}, or
a ``k-essence'' 
scalar field with non-canonical kinetic terms in the Lagrangian ~\cite{Armendariz-Picon:1999rj}-\cite{Chiba:1999ka}, and
string-inspired models such as the contribution of nonlinear short distance physics to vacuum
energy~\cite{transplanck}, and modified Friedman equations at late
time~\cite{cardassian} or large distances~\cite{dgp}.

Dark energy can also receive contributions from topological defects
produced at phase transitions in the early universe (see e.g.
\cite{vilenkin}).

However, despite a well established
theoretical framework, topological defects have not been thoroughly 
explored due to technical difficulties in the numerical simulations.
Moreover, cosmic fluids with $w_X <0$ have an imaginary sound speed
$c_s$ which causes diverging instabilities on small scales
incompatible with structure formation.

More recently, a plausible version of dark energy
made of a frustrated network of domain walls
was proposed by (\cite{battye}, \cite{bucher}).
In these ``solid dark matter'' models (see also \cite{eichler}), 
a negative equation
of state can avoid the short-length instabilities by an elastic
resistance to pure shear deformations. Structure formation is
therefore preserved and CMB anisotropies are affected only
on very large angular scales ($\ell \le 20$ \cite{battye}).

These models have several appealing features:
Firstly, domain walls are ubiquitous in field theory
and unavoidable in models with spontaneously broken symmetries.
Second, the scale of spontaneous symmetry breaking responsible for
the walls is expected to lie in the $10-100$ $KeV$ range
and can arise naturally in supersymmetric theories (\cite{friedland}).
In this respect, the domain wall models of dark energy seem much more
natural than the quintessence models
which assume the existence of a scalar field with a mass of order $10^{-33}
eV$. Finally, two firm phenomenological predictions can be
made for domain walls models: an equation of state strictly
$-1/3 \ge w_X \ge -2/3$ (\cite{friedland}) and a sound speed which can
be a fraction of the speed of light i.e. $c_s \le 1$ (\cite{battye}).
These models are therefore predictive in the value of the equation
of state parameter and distinguishable from
a cosmological constant even at zero order on $w_X$, (while, for example, 
scalar field models can also produce $w_X\sim-1$ although they differ from 
a cosmological constant which in the first order variation has 
$\dot w_X = 0$).

However, recent combined analyses of CMB,
galaxy clustering and SN-Ia luminosity distances data,
have constrained $w_X < -0.82$ at
$95 \%$ C.L. (\cite{spergel},\cite{mmot},\cite{tegmark2}) and therefore
seem to  rule out domain walls.
It is important to notice that the upper bounds on $w_X$ were obtained
under the assumption of a specific choice of priors namely the
popular values for the cosmological parameters in agreement with the
concordance standard model. Therefore the following questions are fully
justified: how model independent are the results of our data analysis and,
are we yet ready to abandon domain walls? In this brief
report we investigate the impact of the priors on the upper bound of $w_X$
by choosing a different data set.
Then we argue that a
different choice of the priors can bring domain walls models in reasonable
agreement with observations.
While the final value of the Hubble constant from the
HST Key Project is $h = 0.72\pm0.02\pm0.07$ (\cite{freedman}), where the
first error is statistical and the second is systematic,
other groups using similar
techniques (see e.g. \cite{saha}, \cite{tammann}, \cite{shanks},
\cite{blanchard}) find a lower value  $h \sim 0.60$.
Measurements based on Sunyaev-Zeldovich method
(see e.g. \cite{birkinshaw} but see also \cite{battistelli}) and on time 
delays for gravitational
lenses (\cite{keeton},\cite{kochanek}) are also suggesting a lower
value $h \sim 0.5$, at least globally.
It is therefore plausible that the true value of $h$ lies in the
lower range allowed by the HST Key project.
This is in contrast with the WMAP constraint $h=0.73 \pm 0.03$
(\cite{spergel}), derived under the assumption of $\Lambda$-CDM.
As we will see, a value of the Hubble parameter $h \le 0.65$ combined
with the WMAP data allows a case to be made for domain walls
models.

Moreover, in the past years, the abundance of high redshift X-ray
selected clusters has been argued to lead to high values of the matter
density parameter $\Omega_m$ (see e.g. \cite{blanchard}).
In particular, analyses of the recent measurements of the
temperature-luminosity relation of distant clusters observed with
XMM-Newton and Chandra satellites, seem to be consistent with
higher values of $\Omega_m \sim 0.8$ (\cite{vauclair}).
Although such high values for a $\Omega_m \simeq 1$ are definitely 
extreme and need to be considered in combination with other data, 
it is conceivable that the true value of
$\Omega_m$ may lie in a range $\Omega_m \sim 0.35 - 0.45$.
Again, this is in tension with the WMAP constraint
$\Omega_m =0.27\pm0.04$ (\cite{spergel})
derived under the assumption of $\Lambda$-CDM.
As we will see in the next section,
domains walls models are clearly accomodated
within the WMAP data when a prior $\Omega_m \ge 0.35$ is
assumed.

Finally, $c_s < 1$,  offers
the advantage of reducing the amplitude of the large-scale CMB anisotropies,
as reported by the WMAP recent data, while
input from the otherwise unknown new physics of the initial conditions of the
Universe is required to bring the standard $\Lambda-CDM$ model
in agreement with the WMAP findings,
~\cite{freeselaura}-\cite{contaldi}.

\section{Analysis}

As is well known (see e.g. ~\cite{Bond:1997wr} and ~\cite{mmot})
a geometrical degeneracy makes virtually impossible any determination of
$\Omega_X$ and $w_X$ from the position of the acoustic peaks in the CMB
anisotropy spectrum. However, if one restrict the analysis to flat
models, a change in $\Omega_X$ must be necessarily compensated by a
change in the matter density $\Omega_m=1-\Omega_X$.
Since for a perfect degeneracy between the CMB peaks one has also
to preserve the physical densities in cold dark matter $\Omega_{cdm}h^2$
and baryons $\Omega_{b}h^2$, the Hubble parameter needs also
to vary.

\begin{figure}[thb]
\begin{center}
\includegraphics[scale=0.28]{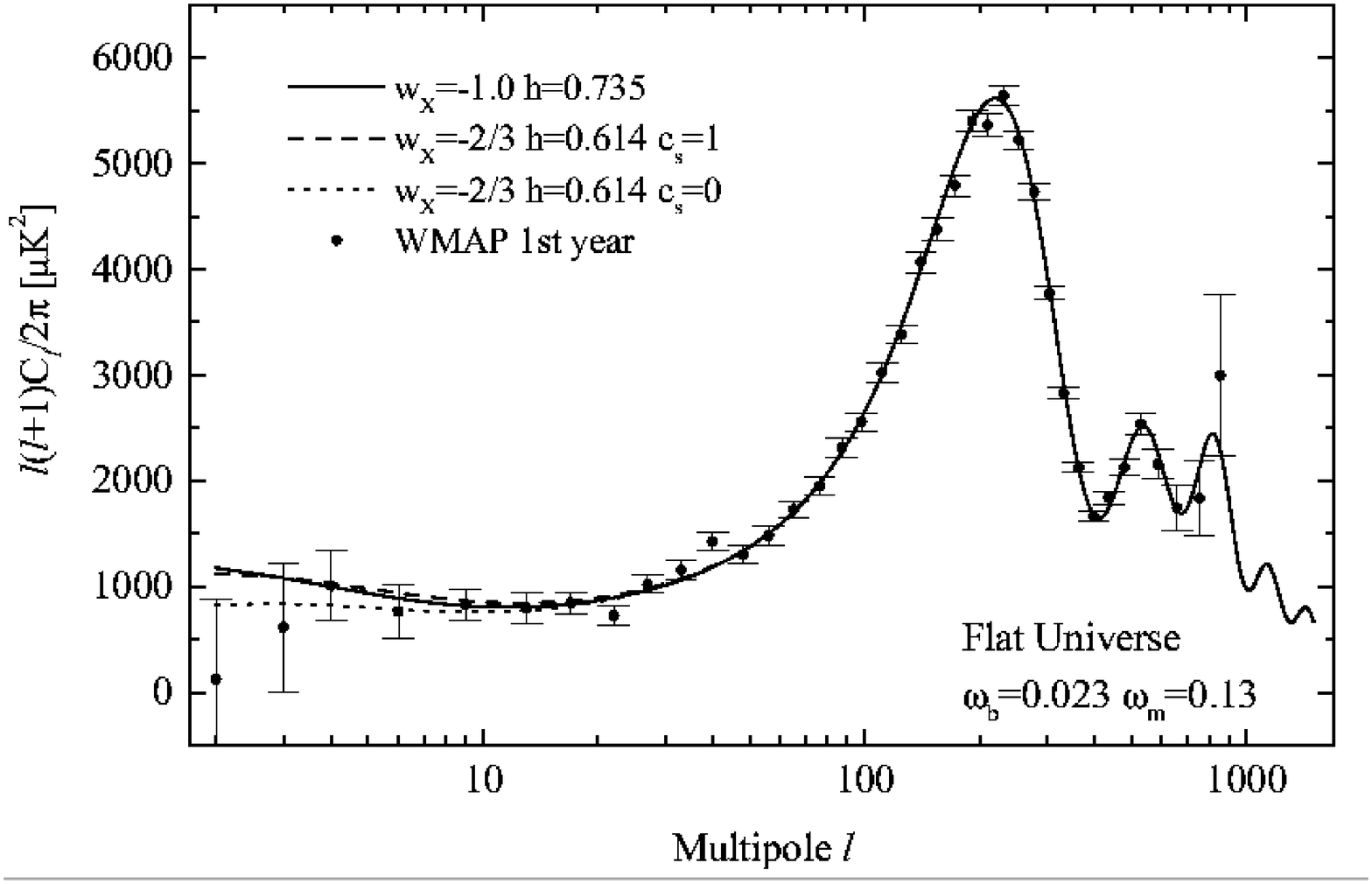}
\includegraphics[scale=0.28]{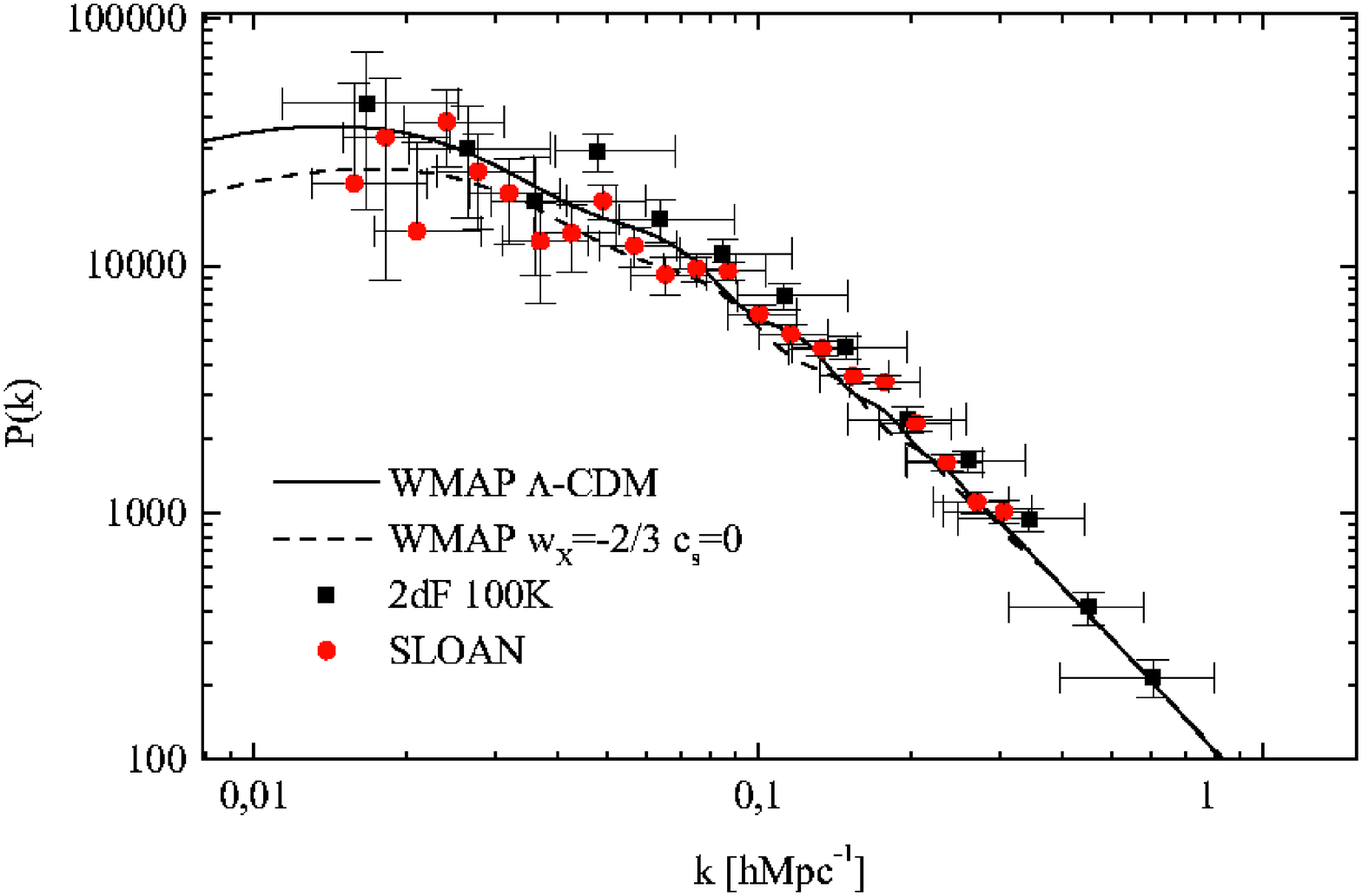}
\end{center}
\caption{Top- Comparison of the $\Lambda$-CDM and Domain Walls
best fit models with 1-st year WMAP CMB data. Bottom- Comparison of the
$\Lambda$-CDM and Domain Walls best fit models with SLOAN and 2dF galaxy
surveys data.}
\label{figomega2}
\end{figure}

\begin{figure}[thb]
\begin{center}
\includegraphics[scale=0.55]{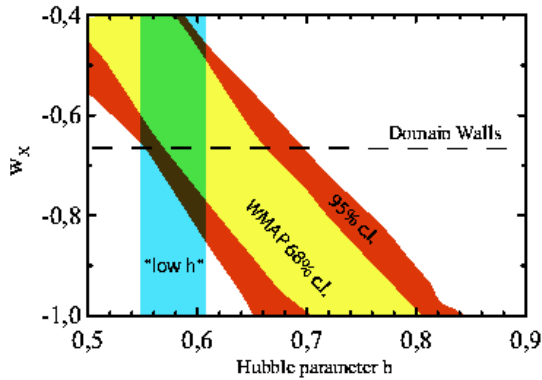}
\end{center}
\caption{$1$ and $2$-$\sigma$ likelihood contours in the $w-h$ plane
from the 1st year WMAP plus ACBAR+CBI data. As we can see, values of the
Hubble parameter $h \sim 0.6$ (shaded region)
are in good agreement with the data and prefers $w \sim -2/3$.}
\label{figomega2}
\end{figure}

\begin{figure}[thb]
\begin{center}
\includegraphics[scale=0.45]{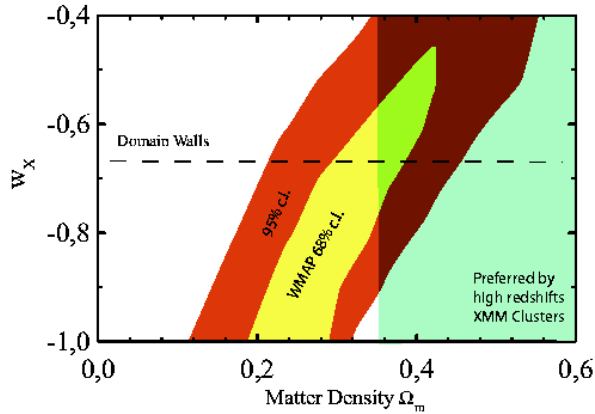}
\end{center}
\caption{$1$ and $2$-$\sigma$ likelihood contours in the $w-\Omega_m$ plane
from the 1st year WMAP plus ACBAR+CBI data. 
As we can see $\Omega_m \ge 0.35$ models.
(shaded region) are excluded at more than $2$-$\sigma$ from the
WMAP+ACBAR+CBI data in case of $w_X=-1$.}
\label{figomega2}
\end{figure}

In Fig.1 (Top Panel) we plot, together with the recent WMAP data, the CMB
temperature power spectrum with parameters $w_X =-2/3$ and $h=0.61$
degenerate with the WMAP $\Lambda$-CDM best fit $w_X=-1$ and $h=0.73$.
The cold dark matter and baryon densities have been
fixed at $\Omega_{cdm}h^2=0.13$ and $\Omega_{b}h^2=0.023$.
Both models have an overall chi-square of $\chi^2 \sim 974$ and are
virtually indistinguishable by the WMAP data.
Also in Fig. 1, bottom panel, we compare the matter power spectra from
best fit CMB domain walls model with the
the real-space power spectrum of galaxies in the 2dF $100$k
and SLOAN galaxy redshift surveys.
Using the data and window functions of the analysis of Tegmark et al.
~\cite{thx} and \cite{tegmark} and marginalizing over
a possible bias $b$ we have found that, on linear scales,
 this model provides a reasonable fit to the present data, and one as
good as $\Lambda$-CDM.

We study the $w_X-h$ degeneracy more quantitatively in Fig.2 where we plot
the WMAP likelihood contours on those $2$ parameters.
The likelihood contours have been computed 
as in \cite{mmot} and include also the ACBAR and CBI datasets.
As we can see, there is a clear degeneracy along the $w_X+h=constant$
direction. Moreover, models with $h \le 0.65$ are excluded at about
$2$-$\sigma$ from the WMAP+ACBAR+CBI data in the case of $\Lambda$-CDM
($w_X=-1$) while models with $h > 0.7$ are excluded
at $2$-$\sigma$ in the case of domain walls
($-2/3 \le w_X \le -1/3$). If one takes at face value
the constraint $h=0.57 \pm 0.03$ from \cite{tammann} this
yields $w_X \ge -0.78$ at $1$-$\sigma$.
We can therefore conclude that while the HST determination
is consistent with $h\sim 0.65$, this is not the case
for the WMAP constraint under the assumption of $\Lambda$-CDM.
An higher value for $w_X$ can solve the discrepancy.

Since the CMB spectrum provides an independent
constraint on $\Omega_{m}h^2$ we
can expect a degeneracy between the equation of state
parameter $w_X$ and $\Omega_{m}$.
We show this in Fig.3 where we plot the
WMAP+ACBAR+CBI likelihood contours on the $w_X-\Omega_m$ plane.
Also plotted in the figure is a region of values compatible
with results from high redshift X-ray clusters. The abundance
of high redshift X-ray selected clusters has been used
to constrain the value of $\Omega_m$ in several works.
The values obtained range from $\Omega_m \sim 1-0.85$
\cite{vauclair}, $\Omega_m \sim 0.85\pm 0.2$ \cite{sadat},
$\Omega_m \sim 0.96 \pm 0.3$ \cite{reichart}.
These results have been obtained under the assumption
of $\Lambda$-CDM. A variation in $w_X$ would affect the growth
factor for these results. However the
effect is small and of a few percent amplitude
(see e.g. \cite{schuecker}). 

We can therefore state conservatively 
that from those high redshift cluster analyses 
$\Omega_m \ge 0.35$. Other high redshift 
cluster analysis suggest a lower value
$\Omega_m \sim 0.35 \pm 0.12$ (\cite{schuecker2}, \cite{borgani})
but are still compatible with the $\Omega_m \sim 0.35-0.45$ 
range.

Again, as we can see, this range is
incompatible at $95 \%$ c.l. with the WMAP constraint obtained under
the assumption of $\Lambda$-CDM. However, higher values of $\Omega_{m}$
are compatible with higher values of $w_X$.
In particular, $\Omega_m \ge 0.35$, indicates $w_X > -0.9$ at
$2$-$\sigma$.

The above results have been obtained under the assumption of a
flat universe, i.e. $\Omega_{Tot}=\Omega_m+\Omega_\Lambda =1$.
While this is certainly one of the most general prediction of inflation,
 is possible to build inflationary models with $\Omega_{Tot}<1$
(see e.g. \cite{linde}). By relaxing the flatness condition,
is possible to obtain degenerate CMB power spectrum by just
decreasing $w_X$ and $\Omega_X$ without modifying the other
parameters (see e.g. \cite{aurich}).
We show this in Fig.4 where a $\Lambda$-CDM flat model is
compared with a degenerate open ($\Omega_{Tot}=0.97$) domain walls model.

\begin{figure}[thb]
\begin{center}
\includegraphics[scale=0.45]{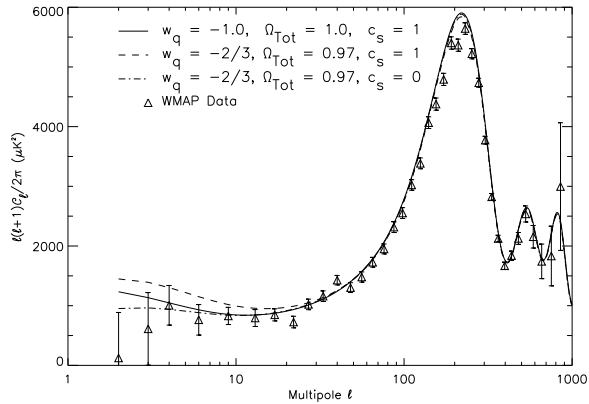}
\end{center}
\caption{CMB flat and open degenerate models.
The parameters have been fixed to $h=0.73$,
$\Omega_mh^2=0.13$, $\Omega_bh^2=0.024$.}
\label{figomega2}
\end{figure}

\begin{figure}[thb]
\begin{center}
\includegraphics[scale=0.3,angle=-90]{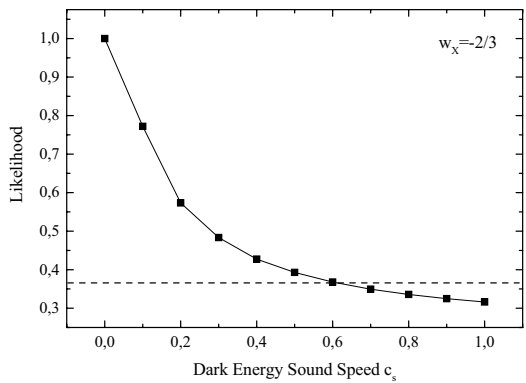}
\end{center}
\caption{Likelihood analysis for the dark energy sound speed
($w_{X}=-2/3$) from the WMAP data. The low quadrupole prefers
$c_s \sim 0$.}
\label{figomega2}
\end{figure}

Another important aspect is the sound speed of the dark energy
component \cite{lewis}.

While for quintessence, the sound speed must be in general equal to
the speed of light, for domain walls models is possible to have
$c_s \le 1$. The main effect of a lower $c_s$ 
is to reduce the power on large scales
(see e.g. \cite{battye}), thus yielding a
better agreement with the low quadrupole as observed by WMAP
(see Fig.1 and Fig.4). Even for quintessence, no perturbations
are expected when $w_X=-1$ (see e.g. \cite{lewis}).

We study the effect more quantitatively in Fig.5 where a
likelihood analysis of the WMAP data is performed by varying
$c_S$ and other cosmological parameters as in
\cite{mmot} but keeping $w_X=-2/3$. In order to simplify the
problem, the perturbations in the ``solid'' dark energy component
are treated as in a non-interacting fluid (\cite{lewis}).
As we can see, a value of $c_s \sim 0$ is preferred by the data, yielding
a weak constraint $c_s \le 0.7$ at $1 -\sigma$ level.

The low CMB quadrupole has raised much interest in recent work
and several physical mechanism have been proposed
to explain this tension (see e.g.\cite{bastero},\cite{contaldi},
\cite{luminet}). The low quadrupole affects the
determination of inflationary parameters ( see e.g.
\cite{spergel},\cite{peiris}, \cite{kinney}) favouring  a
non-zero running or scale-dependence $dn_S/dlnk\sim-0.04$
of the scalar perturbations spectral index $n_S$.
Simple inflationary models predict a running which
is an order of magnitude lower. It is therefore important
to address the question whether a different modelling
of the dark energy component can explain the low
quadrupole. We do this in Fig.6, where we plot the WMAP likelihood
curves for the running for the cases $c_S=1$ and $c_S=0$.
All other parameters are fixed as in the case $w=-2/3$ above,
except the spectral index $n_S=0.93$ as in the reported WMAP running index
best fit \cite{spergel}. As we can see, lowering $c_S$ shifts the
likelihood to higher values of $dn_S/dlnk$, yielding models
with zero running in better agreement with the data.

\begin{figure}[thb]
\begin{center}
\includegraphics[scale=0.3,angle=-90]{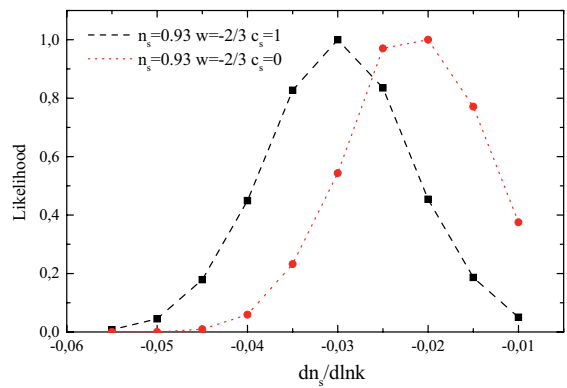}
\end{center}
\caption{Likelihood analysis for the running of the spectral
index $dn_S/dlnk$ from the WMAP data. $c_s \sim 0$ shifts the
likelihood towards $dn_S/dlnk \sim 0$.}
\label{figomega2}
\end{figure}

\section{Conclusions}

Recent combined analysis of cosmic microwave background, galaxy clustering
and supernovae type Ia data have set strong constraints on the
equation of state parameter $w_X$. The bound $w_X < -0.82$ at $95 \%$ c.l.
rules out an important class of models as those based
on domain walls ($-1/3 > w > -2/3$).
Here we have investigated the stability of this result under
a different choice of datasets and theoretical modelling.
Our conclusion is that domain walls
models are not ruled out by the
data and in agreement with the WMAP findings when priors for a ``low''
hubble parameter ($h \le 0.65$),
or for a ``high'' matter density ($\Omega_m \ge 0.35$)
are assumed. Those priors are compatible with
most of current cosmological observations and motivated
by several others.
Moreover, if one relaxes the flatness condition,
it is possible to construct CMB spectra for domain walls
degenerate with $\Lambda$-CDM models that keep the same values of the
physical parameters $h$, $\Omega_m$, $\Omega_b$.
The current CMB evidence for a flat universe therefore relies
on the assumption of the cosmological constant as the dark energy
component. A different value of the sound speed $c_S$ can also lead to
a biased determination of inflationary parameters such as
the running $dn_S/dlnk$.
When compared with CMB data, the domain walls are compatible but
not favoured by the HST constraint on the Hubble parameter.
A value of $\Omega_m =0.35$ and $w_X=-2/3$ is not preferred
by present SN-Ia data (see e.g. \cite{tonry}, \cite{knop}).
When compared with the fiducial WMAP
$\Lambda$-CDM best fit the disagreement is 
$\chi^2_{w_X=-2/3} - \chi^2_{\Lambda} \sim 4.1$
i.e. a $\sim 2.1 \sigma$ disagreement. The SN-Ia dataset has therefore
the biggest weight in ruling out domain walls models in recent combined 
analysis. The latest SN-Ia results seems also to favour 
models with $w_X <-1$ or Chaplygin gases (see e.g. \cite{alam}).
Density profiles of dark matter halos seem also to prefer
$w_X<-1$ \cite{kuhlen} while statistics 
of giant arcs in galaxy clusters prefer $w_X\sim-2/3$\cite{bartelmann}.
A value of $\Omega_m\sim0.35$ is  strongly ruled out in the 
the cluster analysis of \cite{bahcall}.

While systematics in SN-Ia might be present (see e.g.\cite{robinson})
future datasets will be able to clearly test ``solid'' dark energy.
As final remark we mention that detection of non gaussianities
in CMB maps (as recently claimed by several groups \cite{vielva}, 
\cite{naselsky}) is a generic prediction of the models
considered here.


\end{document}